# Selective control of oxygen sublattice stability by epitaxial strain in Ruddlesden-Popper films

*Tricia L. Meyer[1], Lu Jiang[1,2], Jaekwang Lee[3], Mina Yoon[3], John W. Freeland[4], Jae Hyuck Jang[1], Dilpuneet S. Aidhy[1], Albina Borisevich[1], Matthew Chisholm[1], Takeshi Egami[1,2,5], Ho Nyung Lee[1]*

[1]Materials Science and Technology Division, Oak Ridge National Laboratory, Oak Ridge, TN, 37831, USA

[2]Department of Physics and Astronomy, University of Tennessee, Knoxville, TN, 37996, USA

[3]Center for Nanophase Materials Sciences, Oak Ridge National Laboratory, Oak Ridge, TN, 37831, USA

[4]Advanced Photon Source, Argonne National Laboratory, Argonne, IL, USA

[5]Department of Materials Science and Engineering, University of Tennessee, Knoxville, TN, 37996, USA

E-mail correspondence: hnlee@ornl.gov

**Oxygen-defect control has long been considered an influential tuning knob for producing various property responses in complex oxide films.**[1–4] **In addition to physical property changes, modification to the lattice structure, specifically lattice expansion, with increasing oxygen vacancy concentrations has been reported often and has become the convention for oxide materials.**[5,6] **However, the current understanding of the lattice behavior in oxygen-deficient films becomes disputable when considering compounds containing different bonding environments or atomic layering. Moreover, tensile strain has recently been discovered to stabilize oxygen vacancies in epitaxial films, which further complicates the interpretation of lattice behavior resulting from their appearance.**[7] **Here, we report on the selective strain control of oxygen vacancy formation and resulting lattice responses in the layered, Ruddlesden-Popper phases, $La_{1.85}Sr_{0.15}CuO_4$. We found that a drastically reduced Gibbs free energy for oxygen vacancy formation near the typical growth temperature for tensile-strained epitaxial LSCO accounts for the large oxygen non-stoichiometry. Additionally, oxygen vacancies form preferentially in the equatorial position of the $CuO_2$ plane, leading to a lattice contraction, rather than the expected expansion, observed with apical oxygen vacancies. Since oxygen stoichiometry plays a key role in determining the physical properties of many complex oxides, the strong strain coupling of oxygen nonstoichiometry and the unusual structural response reported here can provide new perspectives and understanding to the structure and property relationships of many other functional oxide materials.**

As witnessed over the last few decades, many functional properties of complex transition metal oxides are often closely related with their oxygen stoichiometry[1,2,4]. This is attributed to the fact that the creation or annihilation of oxygen vacancies can modify the valence state of a transition metal or change the charge carrier concentration, and hence modify material's electronic, magnetic, or ionic properties.[8] In addition to changes to the electronic structure, it is well-established that strong coupling exists between oxygen vacancy formation and the lattice structure.[5] In particular, lattice expansion is the most commonly reported observation resulting from an increase in oxygen defects.[5–7,9] However, it is important to mention that exceptions to this include oxygen-deficient perovskite, $SrTiO_{3-\delta}$, which exhibits little to no lattice change[10–13] and doped-$CeO_{2-\delta}$[14,15], which exhibits lattice contraction upon vacancy formation due to the smaller size of the oxygen vacancy compared to the $O^{2-}$ ion. Thus, this lattice behavior relies heavily on the cation and anion environments and is not universal for all complex oxides, though this is often the assumption.

Recently, first principle calculations on a simple perovskite suggested that strain is a good tool to lower the oxygen vacancy formation energy, though the energy change observed was not drastic (<10% by 2% strain).[7] Interestingly, these results indicated that tensile strain was more favorable for producing oxygen defects, mostly due to a change in the crystal structure (i.e. increased octahedral tilting). Abiding by Poisson's ratio for volume conservation, a material under tensile (compressive) strain will have a contracted (expanded) out-of-plane lattice parameter ($c$) as compared to the respective bulk value. Thus, deconvoluting the structural response resulting from epitaxial strain and oxygen vacancy formation in strained films poses new challenges. In the case of open-framework structures or layered RP phases such as superconducting $La_{2-x}Sr_xCuO_4$, strained films of this structure-type have potential for losing or gaining oxygen even more easily than simple perovskites.[16–18] Such redox sensitivity is a promising application, in particular, for oxides in energy generation and storage.[19–22] However, the precise understanding necessary for realizing this potential, particularly regarding the interplay among epitaxial strain, lattice structure and oxygen defects, is lacking even after decades of research.

In this letter, we report on the structural behavior of strained, $La_{1.85}Sr_{0.15}CuO_4$ (LSCO) films in the presence of oxygen vacancies. In agreement with earlier reports, we find that tensile-strained films inherently form oxygen vacancies, whereas compressively strained films exhibit a robust, oxygen stoichiometric structure. However, oxygen vacancies in our LSCO films preferentially form in the equatorial oxygen site of the $CuO_2$ plane. More importantly, we found through extensive experimental and theoretical studies that site-specific oxygen vacancy formation leads to two distinct lattice responses: (1) contraction upon oxygen vacancy formation within the equatorial site and (2) expansion when the vacancy is in the apical site. Thus, it is imperative to amend the convention that oxygen vacancy formation leads only to chemical expansion, since much of the understanding of structure-property relationships in complex oxide films has mainly relied on this postulation.

Since strain plays a large role in controlling the oxygen defect concentration[7], we have grown a series of epitaxial LSCO ($a$ = 3.779 Å, $c$ = 13.226 Å)[23] films using pulsed laser epitaxy on $LaSrAlO_4$ (LSAO, $a$ = 3.756 Å) and $LaAlO_3$ (LAO, $a$ = 3.788 Å) single crystalline substrates to induce both compressive (−0.61%) and tensile (0.23%) strain within the cuprate lattice, respectively. In order to isolate the influences of strain from changes in oxygen content in the Sr-doped RP cuprates, we exposed each set of films to highly reducing and oxidizing atmospheres at temperatures as high as 500 °C. These annealing conditions were chosen since the oxygen incorporation should increase with increasing annealing temperature, thus providing systematic control of the oxygen stoichiometry within the set of films. We confirmed that the in-plane lattice constants of these ultrathin (20 nm in thickness) films are coherent with those of the substrates based on x-ray reciprocal space mapping (see Supplementary Fig. S1). Scanning transmission electron microscopy investigation confirmed the growth of high quality thin films with well-defined interfaces (see Supplementary Fig. S2).

Under oxidizing conditions, there are three possibilities of oxidation: (1) filling oxygen vacancies that are created during the high temperature growth; (2) inserting interstitial oxygen in stoichiometric as-grown films; and (3) no changes due to the high thermodynamic energy required for oxidation or reduction. In Figure 1, we show the $c$ lattice parameter response as a function of different post-annealing conditions, i.e. oxygen partial pressure, in order to distinguish the mechanism of oxidation in our LSCO films. The out-of-plane lattice constants of oxidized, tensile-strained films are expanded by as much as 0.3%, whereas the reduced film by annealing in vacuum showed a 0.2% contraction as compared to the as-grown film. This is surprising since the redox-lattice response is opposite to the convention that oxygen vacancies lead to chemical expansion.[5] The even more surprising behavior is that compressive-strained films exhibit no obvious changes in $c$, revealing persistently strong structural stability under a variety of redox conditions. Since the lattice parameters of oxides are highly sensitive to oxygen non-stoichiometry, and since the films were grown under identical conditions, we attribute such variation to strain-driven changes in the oxygen vacancy concentration. If we compare the as-grown films for both strained states, we see that the out-of-plane lattice parameter for the tensile-strained as-grown film is significantly smaller than both bulk and compressively-strained film values. This result has not been observed for other $La_{2-x}Sr_xCuO_4$ thin films when grown under extremely oxidizing conditions or in activated oxygen[17,18,24], which implies that the mechanism of oxidation is different under these conditions.

In order to elucidate this unusual lattice response, we employed first-principles density functional theory (DFT) calculations to explore the energetics of vacancy formation in LSCO under different biaxial strains. The oxygen vacancy formation energies ($\Delta E^{Vo}$) of both apical (perpendicular to the $CuO_2$ plane) and equatorial (in plane of the $CuO_2$ plane) oxygen vacancies (see schematic in Fig. 2a) are presented in Fig. 2b. Their relativestabilities are also compared as a function of strain. This result illustrates that the equatorial vacancy is energetically more favorable than the apical vacancy by nearly 30%. Furthermore, compressive strain increases the vacancy formation energy for both vacancy sites with respect to the unstrained case, whereas tensile strain shows no such increase. The strongly contrasting oxygen sublattice behaviors—remarkably high stability of oxygen in compressively strained films and promoted formation of oxygen vacancies in tensile-strained films—consistently supports our experimental observations. It is worth mentioning that a study on the isostructural cathode material $La_2NiO_4$ has shown a similar stability of equatorial oxygen vacancies.[31]

It is evident from Fig. 2b that the energy scale for $\Delta E^{Vo}$ is unrealistically large to reach experimentally, as can also be seen in previous reports.[7] In order to accurately evaluate the strain-dependent oxygen vacancy formation behavior in LSCO, we also considered the thermodynamics as functions of temperature and pressure for oxygen vacancy formation by calculating the Gibbs free energy [$\Delta G^{Vo}(T)$] of the system. We focused on the equatorial oxygen defect formation as this pertains to our films and compared changes in the $\Delta G^{Vo}(T)$ before and after creating an oxygen vacancy as a function of strain. The chemical potential of molecular oxygen in the gas phase and the vibrational free energies ($\Delta F_{vib}$) are taken into account for a wide range of oxygen pressure ($10^{-6}$ – 760 Torr, see Supplementary Fig. S3). As shown in Fig. 2c, the $\Delta G^{Vo}(T)$ required for creating equatorial oxygen vacancies in a tensile-strained film can be significantly reduced when including $\Delta F_{vib}$. It is important to mention that a reduction of the $\Delta G^{Vo}(T)$ of ~90% for 1.4% tensile strain occurs at the growth environment of 700 °C and 100 mTorr $O_2$. Also, the $\Delta G^{Vo}(T)$ either increases or decreases when exposed to oxidizing (760 Torr) or reducing conditions ($10^{-6}$ Torr), respectively. In agreement with our experimental data, $\Delta G^{Vo}(T)$ for compressive-strained films is much less sensitive to the oxygen pressure change. These data clearly illustrate the reason for oxygen loss in tensile-strained films during the growth and explain the strong stability of the compressive-strained films to reducing conditions.

Based on the thermodynamic calculations and X-ray diffraction data, one can conclude that the reduction of the unit cell volume in tensile-strained films originates from the preferential formation of oxygen vacancies. In order to understand this unexpected phenomenon – lattice shrinkage upon vacancy generation – we have calculated the lattice constant change ($\Delta c = c_{vacancy} - c_{no\ vacancy}$) by creating an oxygen vacancy as a function of strain and vacancy position (see Fig. 3). Regardless of the type of strain, creating oxygen vacancies at the equatorial position consistently results in a contraction of the $c$ lattice parameter as compared to the bulk value, in good agreement with our experimental observation in Fig. 1. In contrast, a $c$ lattice expansion is predicted when oxygen vacancies are created instead at the apical position. The predicted lattice responses to the vacancy formation clearly support the preferential formation of equatorial oxygen defects in tensile-strained LSCO, which has not been previously considered.

The redox behavior of strained cuprates suggests that these could be candidate materials for various electrochemical and energy applications. Provided that oxygen vacancy stability in LSCO is reliant upon the strain state, it is quite likely that the diffusion of oxygen should exhibit a similar trend since ionic motion tends to be facilitated by oxygen vacancies. Therefore, we have investigated the oxygen activation energy, which is the energy required for an oxygen atom to diffuse from one site to the other, as a function of strain in order to gain more insight into the motion of oxygen defects. Remarkably, as shown in Figure 4a, we have found that 1.4% tensile strain greatly facilitates oxygen migration by reducing the energy barrier (Fig. 4b) by almost 20% from the unstrained case. Conversely, the same amount of compressive strain restricts migration due to the increased energy barrier. Overall, it is clear that tensile strain facilitates not only the formation of oxygen vacancies in LSCO, but also oxygen ion motion within the material. Open framework RP phases (*e.g*., our LSCO) were recently spotlighted for their potential for electrochemical applications[19–21,25]. Thus, our observation of the improved oxygen activities in tensile-strained thin films provides crucial information for functionalizing oxygen defects for these applications.

In summary, the strong coupling between epitaxial strain and the oxygen vacancy concentration in the RP phase LSCO is reported. A highly contrasting effect on oxygen sublattice stability, i.e. tensile strain facilitating oxygen vacancy generation, whereas compressive strain enhances the oxygen stability, has been both experientially and theoretically discovered. Unlike simple perovskites, we have found that the lattice volume of the RP phase, LSCO, contracts upon generation of oxygen vacancies due to preferential formation of oxygen vacancies at the equatorial site. We also note that the formation of apical oxygen vacancies can expand the lattice volume, but our computation results indicate that it is energetically unfavorable. This new understanding of oxygen defects in systems with more complex structural configurations provides a link to discriminating between different structure and property relationships. Additionally, recognizing that compressive strains as small as 0.61% can greatly increase the stability of the oxygen sublattice, even in a highly reducing environment, has large implications for many oxide electronic devices made of functional oxides, in which oxygen non-stoichiometry can be detrimental to the functionalities, such as ferroelectricity, magnetism, and superconductivity. On the contrary, our discovery that tensile strain promotes formation of oxygen vacancies in a variety of different atmospheres and at low temperatures has huge implications for developing oxide-based catalysts for low or medium temperature solid oxide fuel cells. This strain engineering approach, thus, provides a new tuning knob to selectively stabilize or destabilize oxygen in layered R-P phase oxides, opening a new door to developing novel materials through functionalization of oxygen defects.

**EXPERIMENTAL SECTION**

**Sample Fabrication.** We have deposited optimally doped LSCO films by pulsed laser epitaxy (PLE). Note that the Sr content is nominally equivalent to the target, though the real film composition may be slightly off due to the highly energetic ablation process. Among various growth conditions, we systematically changed the lattice constant of substrates in order to change both the sign of strain (i.e. either tensile and compressive) and degree of the strain by using various perovskite-based substrates as stated in the main text. LSCO films were grown in an oxygen atmosphere of 100 mTorr of $O_2$ at a growth temperature of 700 °C. For the as-grown samples, films were quenched in 100 Torr $O_2$ and cooled to room temperature. For the post-annealing protocols (vacuum and $O_2$ annealing), samples were cooled to 500 °C and then post annealed for 30 minutes. The samples were cooled to room temperature in the atmosphere used for post-annealing. The laser energy was fixed to 0.5 J/cm$^2$.

**Structure.** Structural data was obtained using a four-circle Panalytical X'pert Pro diffractometer. Representative XRD spectra, including $\theta-2\theta$ scans and reciprocal space mapping of the high-quality LSCO films discussed in the text are illustrated in Supplementary Fig. S1.

**Electron Microscopy.** Images were recorded using an aberration-corrected scanning transmission electron microscope (Nion UltrasSTEM at 200 kV).

**Theoretical Calculations.** Calculations were performed using density functional theory (DFT) in the generalized gradient approximation (GGA) and the projector-augmented wave method with a plane-wave basis as implemented in the Vienna ab initio simulation package (VASP) code. The large size of 2×2×1 supercell ($La_{16}Cu_{32}O_{64}$) is considered to avoid spurious interactions between oxygen vacancies and Sr dopants. A Hubbard U correction with $U_{eff}$ (U-J) =7.0 eV was applied to improve the description of the Cu-3$d$ states. We consider a biaxial strain in the $ab$-plane to explore the lattice constant change and identify the preferential site for oxygen vacancy formation in $La_{2-x}Sr_xCuO_4$ thin films by using the defect structure $Sr_1La_{15}Cu_{32}O_{63}$, containing 111 atoms. The Gibb's Free energy change was calculated by taking into consideration $\Delta F_{vib}$ as a function of both strain and temperature. The frequency dependent vibrational mode contributions to $\Delta F_{vib}$ for the structure containing vacancies (defective cuprate) and those with no vacancies (pure), as shown in Supplementary Fig. S4. Further details regarding the calculation method and parameters can be found in the Supplementary Information.

## ACKNOWLEDGEMENTS

We would like to gratefully acknowledge D. G. Schlom and S. Park for the fruitful discussions. This work was supported by the U.S. Department of Energy (DOE), Office of Science, Basic Energy Sciences (BES), Materials Sciences and Engineering Division. The theoretical calculations were performed as a user project at the Center for Nanophase Materials Sciences, which is sponsored at Oak Ridge National Laboratory by the Scientific User Facilities Division, BES, U.S. DOE. The use of the Advanced Photon Source was supported by the U.S. DOE, BES, under Contract No. DE-AC02-06CH11357.

## AUTHOR CONTRIBUTIONS

T.L.M. and L.J. grew samples and collected the structural data. J. H., A.B. and M.C. collected and analyzed STEM images. J.L. and M.Y. performed the theoretical calculations. J.W.F. and D.A. participated in discussions and data analysis. T.L.M. and H.N.L. wrote the manuscript with discussions and improvements from all authors. H.N.L. initiated and coordinated the research.

**FIGURES**

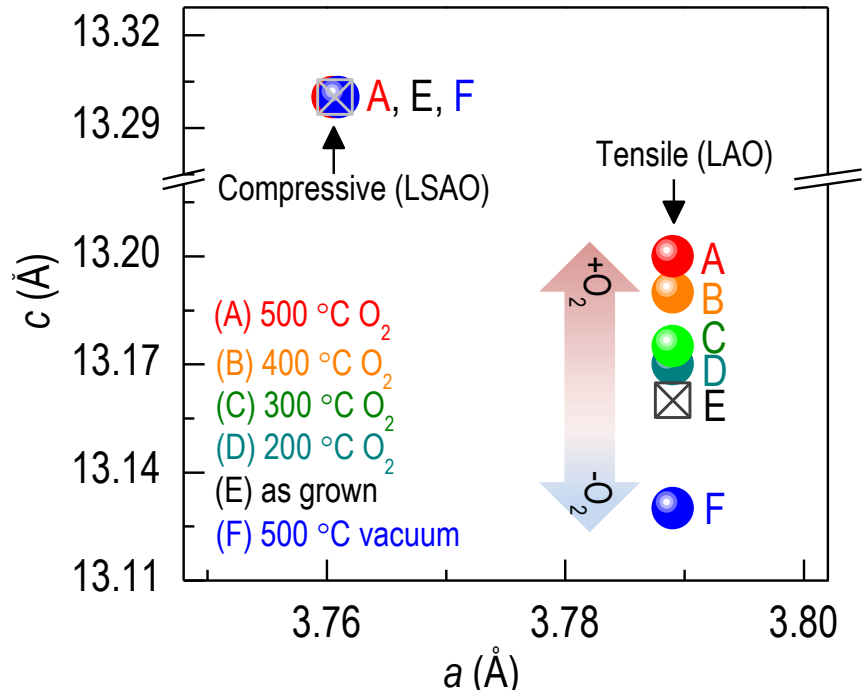

**Figure 1. Strain dependent stability of lattice.** Out-of-plane $c$ lattice constants of 20-nm-thick LSCO films on LAO and LSAO substrates. The lattice constants were measured from as-grown, oxygen and vacuum annealed films annealed at temperatures from 200 to 500 °C. Tensile-strained films on LAO indicate a lattice contraction (expansion) upon reduction (oxidation), unlike compressively strained films on LSAO that show no change in lattice under identical conditions. The in-plane lattice is fixed with the substrate in all films, as confirmed by x-ray reciprocal space mapping.

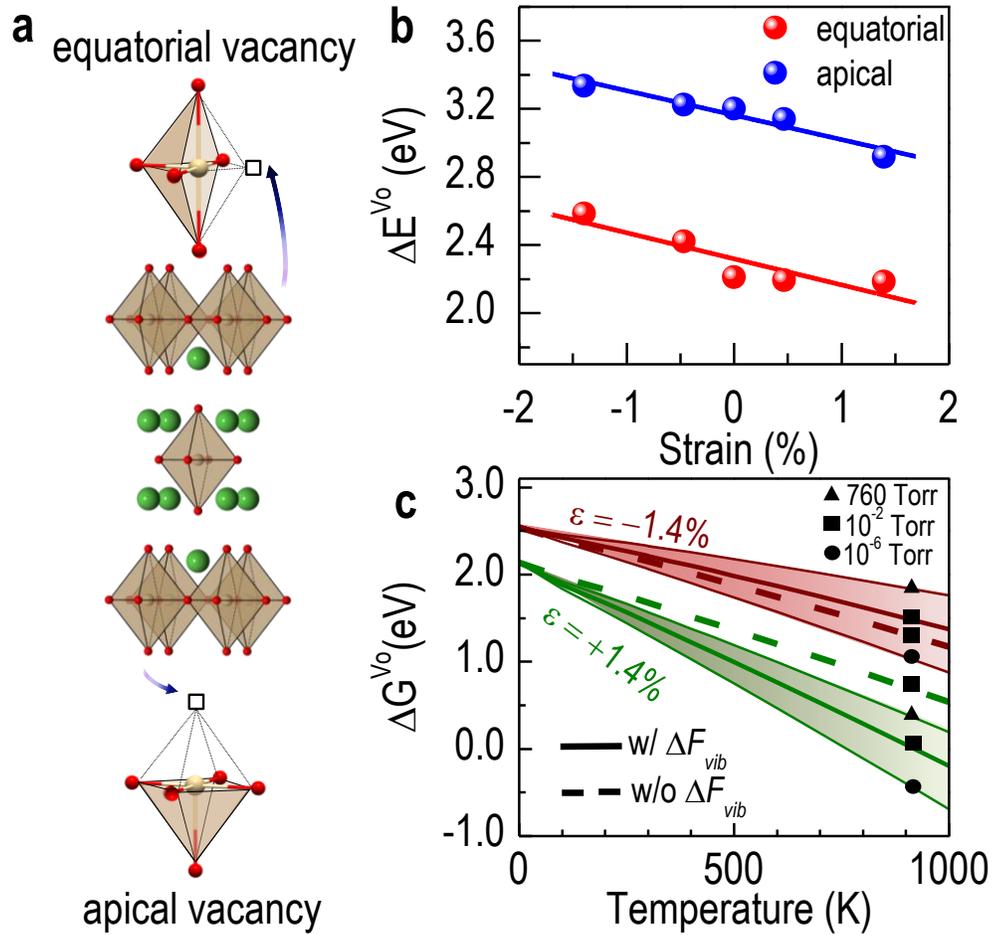

**Figure 2. Energetics of oxygen vacancy formation**. **a**, Schematic of LSCO structure illustrating the two possible oxygen vacancy sites. **b**, Strain dependent oxygen vacancy formation energy of equatorial and apical oxygen vacancies. The formation of oxygen vacancies at equatorial sites requires ~30% less energy than apical sites. **c**, Gibbs free energy as a function of temperature for the formation of equatorial oxygen vacancies in both compressive- (top) tensile-strained (bottom) LSCO films, with (solid lines) and without (dashed lines) lattice vibrations, $\Delta F_{vib}$. Three different pressures are considered: 760 Torr (maroon), $10^{-2}$ Torr (white) and $10^{-6}$ Torr (green) to verify the sensitivity and stability of strained films in different atmospheres.

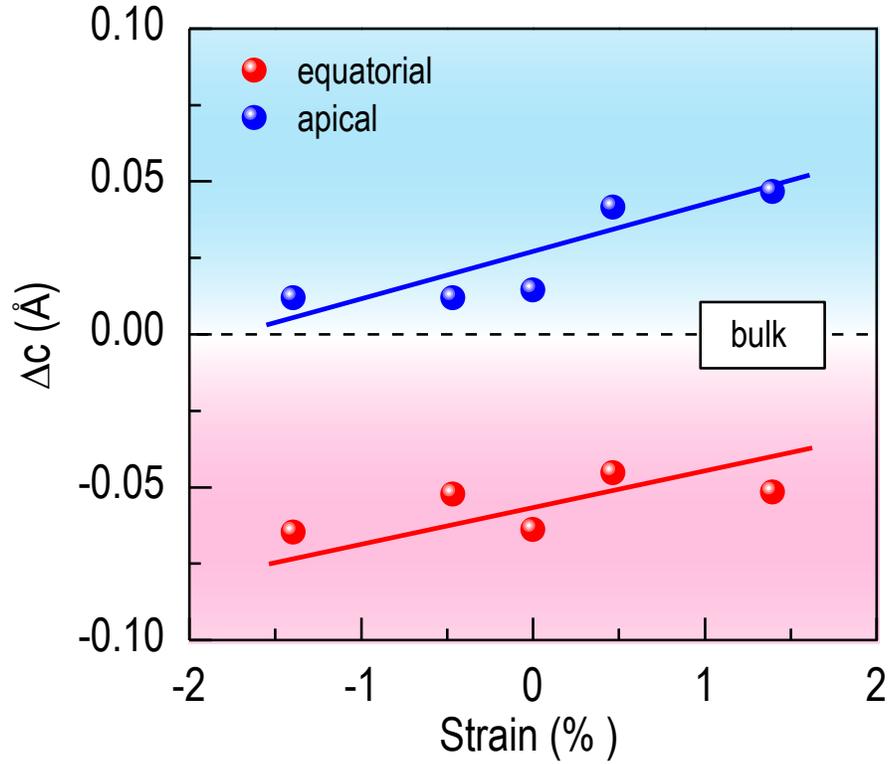

**Figure 3. Contrasting strain dependent lattice volume change upon formation of oxygen vacancies**. Strain-dependent $c$ lattice constant change ($\Delta c = c_{\text{vacancy}} - c_{\text{no vacancy}}$) for the cases of oxygen vacancies created in either equatorial (red symbols) or apical (blue symbols) sites. This site-specific lattice response confirms that our experimental observation originates from the preferential oxygen vacancy generation in the equatorial site.

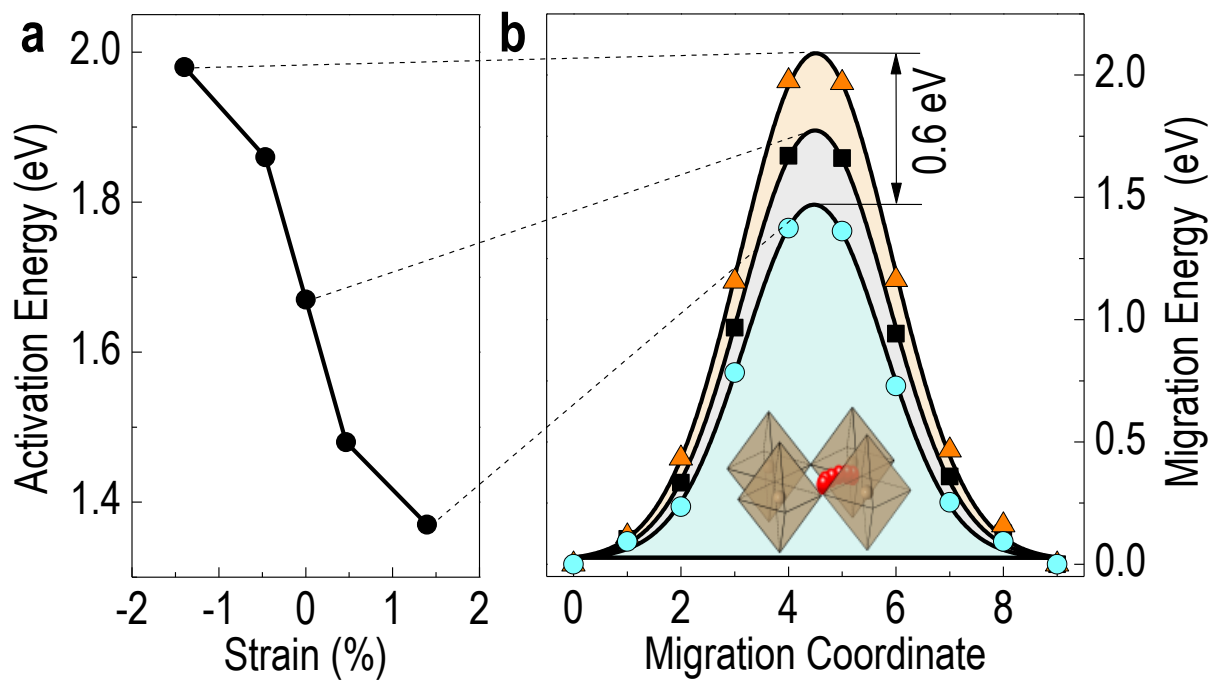

**Figure 4. Stabilization of oxygen by compressive strain and promoted oxygen migration under tensile strain**. **a**, Activation energy barrier for oxygen movement as a function strain. The path between two equatorial oxygen sites was considered due to the preferential oxygen vacancy formation along this position as schematically shown in the inset. **b**, migration energy barrier associated with the −1.4, 0 and 1.4% strain, as designated in **a**.

# SUPPORTING INFORMATION

**Supplementary Figures.**

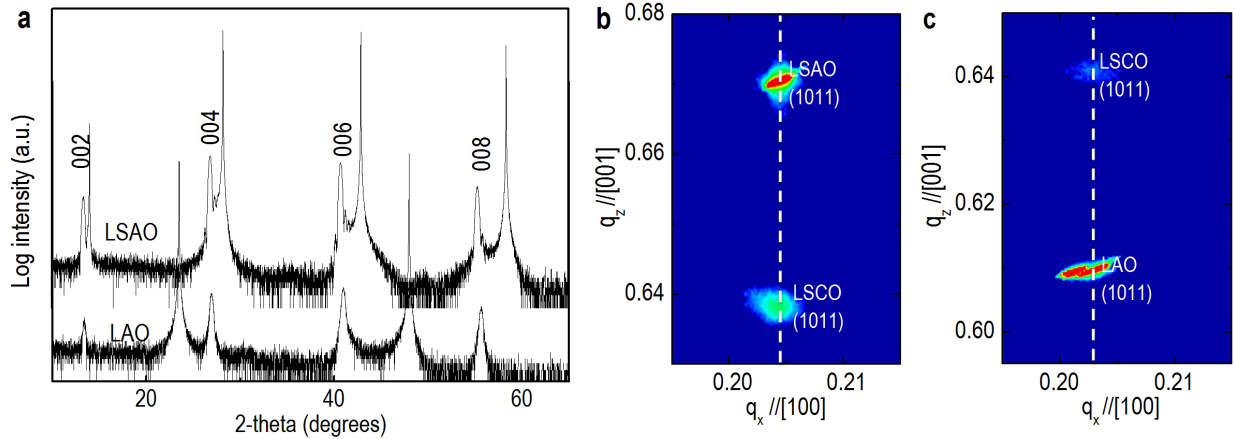

**Figure S1. Structural characteristics of epitaxial La$_{1.85}$Sr$_{0.15}$CuO$_4$ films.** X-ray diffraction spectra of LSCO films grown on LSAO ($a$ = 3.756 Å) and LAO ($a$ = 3.79 Å) single crystal substrates. The (00$l$) peaks of LSCO film on each substrate show that the LSCO is grown along the (001) direction and confirms the high quality of the films and absence of secondary phases. The reciprocal space mapping of the (1 0 $\overline{11}$) peak of LSCO films grown on each substrate are showed below. The indexes of the peaks are labeled. Films grown on LAO and LSAO are fully strained while films on STO and LSAT exhibit strain relaxation.

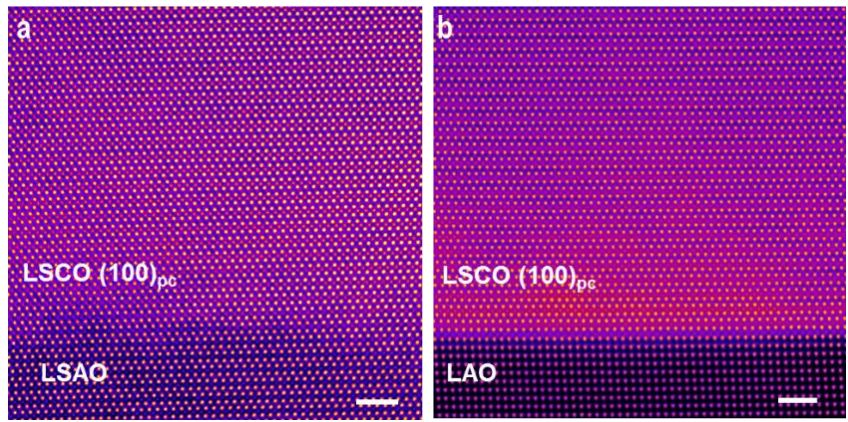

**Figure S2. Transmission electron microscopy images of strained LSCO films. a, b,** HAADF image of LSCO on LSAO and LAO substrates along (100)$_{pc}$ direction. LSCO films are well grown epitaxially on both substrates. The scale bar in both figures is 2 nm.

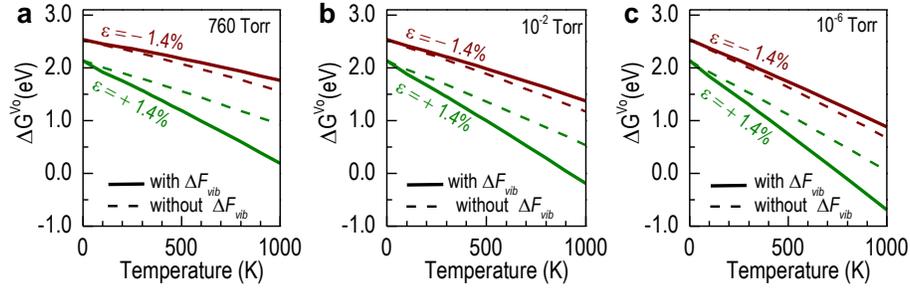

**Figure S3.** Gibb's free energy as a function of strain at pressure of **a**, 760 Torr, **b**, $10^{-2}$ mTorr, **c**, $10^{-6}$ Torr. The energy has been considered with (solid line) and without (dashed line) the vibrational free energy contributions, $\Delta F_{vib}$. The data show that vacancy formation is thermodynamically more stable when films are tensile-strained than when under compressive strain. For all pressures, compressive strained films have higher Gibbs free energy when the vibrational contributions are considered, which supports the robust lattice observed in Figure 3 of the main text in even reducing (vacuum annealing) conditions. It is abundantly clear that reducing conditions (**c**) favor oxygen vacancy formation in tensile strained films, as shown by the small Gibbs Free energy.

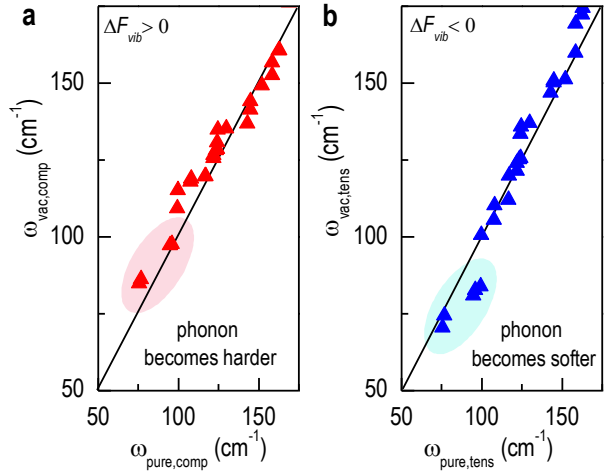

**Figure S4. Comparison of phonon frequencies of defective and pure LSCO.** Comparison of phonon frequencies ($\omega_{vac}$) of defective (with vacancies) and pure ($\omega_{pure}$) LSCO under **a**, −1.4% compressive strain **b**, 1.4% tensile strain. Here, the phonon mode hardens for compressive-strained LSCO, as indicated by the positive $\Delta F_{vib}$, whereas tensile strain softens the phonon, as indicated by the negative $\Delta F_{vib}$.

## Supplementary Methods.

*Theoretical*

Calculations were performed using density functional theory (DFT) in the generalized gradient approximation (GGA) and the projector-augmented wave method with a plane-wave basis as implemented in the Vienna ab initio simulation package (VASP) code. The large size of 2×2×1 supercell ($La_{15}Sr_1Cu_{32}O_{64}$) is considered to avoid spurious interactions between oxygen vacancies and Sr dopants. A Hubbard U correction with $U_{eff}$ (U-J) =7.0 eV was applied to improve the description of the Cu-3$d$ states.

We consider a biaxial strain in the *ab*-plane to explore the lattice constant change and identify the preferential site for oxygen vacancy formation in $La_{2-x}Sr_xCuO_4$ thin films. Such a biaxial strain is imposed by altering the simulation cell in the *ab* direction and relaxing the cell dimension in the *c*-direction. For the defect structure, one oxygen atom was removed in equatorial and apical site, respectively. The resultant defect structure ($La_{15}Sr_1Cu_{32}O_{63}$), containing 111 atoms, was relaxed until the force in all the ions was smaller than 50 meV/Å per ion using a 2×2×2 k-point mesh centered at the Γ- point, with an energy cutoff of 400 eV. The formation energy ($\Delta E^{Vo}$) of the oxygen vacancy was calculated using the following equation:

$$\Delta E^{Vo} = (E^{Vo} - E^p) + 1/2 E(O_2), \quad (S1)$$

where superscripts *Vo* stands for a defect structure ($La_{15}Sr_1Cu_{32}O_{63}$) containing a Sr substitutional atom and an oxygen vacancy, and *p* stands for a perfect structure ($La_{15}Sr_1Cu_{32}O_{64}$) containing only a Sr substitutional atom, respectively. $E^{Vo}$ is the total energy of a defect structure, $E^p$ is total energy of a perfect structure and $E(O_2)$ is the total energy of diatomic oxygen molecule.

The difference in Gibbs free energy is given by neglecting the pressure-volume (*PV*) term, which is in the order of ~$10^{-6}$ eV in our case:

$$\Delta G^{Vo}(T) = \Delta E^{Vo} + \left[(E^{Vo}_{vib} - TS^{Vo}_{vib}) - (E^p_{vib} - TS^p_{vib}) + 1/2\Delta\mu_{o2}\right], \quad (S2)$$

where $E_{vib}$ is the vibrational contribution to internal energy, $S_{vib}$ is vibrational entropy, and $1/2\Delta\mu_{o2}$ is the vibrational contribution to the chemical potential of $O_2$ in the gas phase. A precise description of the change in the vibrational free energy of the solids, $\Delta F_{vib}$, requires computationally expensive calculations of the full phonon spectrum of perfect and defective LSCO, while $1/2\Delta\mu_{o2}$ can be taken from the thermodynamic database[1]. Using harmonic approximation the free energy change due to lattice vibration is described by

$$\Delta F_{vib} = (E^{Vo}_{vib} - TS^{Vo}_{vib}) - (E^p_{vib} - TS^p_{vib}) = k_B T \int_0^\infty \Delta D(\omega) \ln\left[2\sinh\left(\frac{\hbar\omega}{2k_B T}\right)\right] d\omega, \quad (S3)$$

where $\Delta D(\omega)$ is the phonon density-of-states difference between defective and perfect LSCO. According to equation S3, the soft vibrational modes with low frequencies dominantly contribute to $\Delta F_{vib}$ at high temperature; thus if the soft vibrational modes in defective LSCO become softer than those in pristine LSCO, then the $\Delta F_{vib}$ will become negative. To qualitatively estimate the contribution of $\Delta F_{vib}$, only zone-centered ($\Gamma$ point) frequencies are calculated by displacing each ion in the direction of each Cartesian coordinate by 0.02 Å for defective and pristine LSCO under compressive (−1.4%) and tensile (1.4%) strain. We find that tensile strain softens the $A_{2u}$ mode with low frequency consisting of the vibration of Cu, along with two planar O atoms, against La atoms[2]; accordingly, $\Delta F_{vib}$ becomes negative (see Supplementary Fig. 4). In contrast, compressive strain hardens the $A_{2u}$ mode, thus increasing $\Delta F_{vib}$. Most of the other phonon modes with high frequencies change randomly with negligible contribution. Thus, we focus on the $A_{2u}$ mode and roughly estimate the maximum changes in $\Delta F_{vib}$ by treating all lattice vibrations as $A_{2u}$ mode for defective ($\hbar\omega_{0,Vo}$) and perfect ($\hbar\omega_{0,p}$) LSCO, given by

$$\Delta D(\omega) = 3N\left[\delta(\omega - \omega_{0,Vo}) - \delta(\omega - \omega_{0,p})\right], \quad (S4)$$

where $\underline{N}$ is the total number of atoms.

1. P.J. Linstrom and W.G. Mallard, Eds., *NIST Chemistry WebBook, NIST Standard Reference Database Number 69*, (National Institute of Standards and Technology, Gaithersburg MD, 20899, http://webbook.nist.gov, retrieved April 1, 2015).

2. Wang, C., Yu, R. & Krakauer, H. First-principles calculations of phonon dispersion and lattice dynamics in $La_2CuO_4$. *Phys. Rev. B* **59,** 9278–9284 (1999)